\documentclass[runningheads]{llncs}
\def\ANON{0}

\usepackage{tikz}
\usepackage{amsmath}
\usepackage{graphicx}
\usepackage{algorithmic}
\usepackage{epstopdf}
\usepackage{rotating}
\usepackage[square,sort,comma,numbers]{natbib}
\usepackage{xurl}
\usepackage{color}
\usepackage{fancybox}
\usepackage{pstricks}
\usepackage{balance}
\usepackage{pst-plot}
\usepackage{array}
\usepackage{multirow}
\usepackage{mathtools}
\usepackage{pstricks-add}
\usepackage{flexisym}
\expandafter\def\expandafter\UrlBreaks\expandafter{\UrlBreaks\do\_}
\usepackage[linesnumberedhidden,titlenotnumbered,ruled]{algorithm2e}
\usepackage{listings}

\usepackage{xcolor}

\definecolor{codegreen}{rgb}{0,0.6,0}
\definecolor{codegray}{rgb}{0.5,0.5,0.5}
\definecolor{codepurple}{rgb}{0.58,0,0.82}

\lstdefinestyle{mystyle}{
	commentstyle=\color{codegreen},
	keywordstyle=\color{magenta},
	stringstyle=\color{codepurple},
	basicstyle=\ttfamily,
	breakatwhitespace=false,         
	breaklines=true,                 
	captionpos=b,                    
	keepspaces=true,                 
	showspaces=false,                
	showstringspaces=false,
	showtabs=false,                  
	tabsize=2
}

\SetKw{Continue}{continue}
\SetKw{Break}{break}
\SetKw{DownTo}{downto}
\usepackage{url}
\usepackage[colorlinks=true,citecolor=purple]{hyperref}
\hypersetup{colorlinks=true}
\newcommand{\MyComment}[1]{$/\ast$ #1 $\ast/$}

\ifnum\ANON=0
\newcommand{\SV}[1]{\textcolor{red}{{\sf (SVNote:}{\sl{#1}}{\sf)}}}
\newcommand{\SM}[1]{\textcolor{blue}{{\sf (SMNote:}{\sl{#1}}{\sf)}}}
\newcommand{\DK}[1]{\textcolor{magenta}{{\sf (DKNote:}{\sl{#1}}{\sf)}}}
\else
\newcommand{\SV}[1]{}
\newcommand{\SM}[1]{}
\newcommand{\DK}[1]{}
\fi

\newcommand{\etal}{\textit{et al. }}

\newenvironment{compactlist}{
	\begin{list}{{$\bullet$}}{
			\setlength\partopsep{0pt}
			\setlength\parskip{0pt}
			\setlength\parsep{0pt}
			\setlength\topsep{0pt}
			\setlength\itemsep{0pt}
			\setlength{\itemindent}{\leftmargin}
			\setlength{\leftmargin}{0pt}
		}
	}{
	\end{list}
}

\newcommand\blfootnote[1]{%
  \begingroup
  \renewcommand\thefootnote{}\footnote{#1}%
  \addtocounter{footnote}{-1}%
  \endgroup
}

\begin{document}

\date{}

\title{Revisiting Driver Anonymity in ORide}

\ifnum\ANON=0

\author{ Deepak Kumaraswamy\inst{1}\orcidID{0000-0003-0245-9520} \and
Shyam Murthy\inst{2}\orcidID{0000-0002-0222-322X}  \and
Srinivas Vivek\inst{3}\orcidID{0000-0002-8426-0859}}
\authorrunning{D. Kumaraswamy et al.}
\institute{National Institute of Technology Karnataka, Mangalore, India \\
	\email{deepakkumaraswamy99@gmail.com} \and	
	International Institute of Information Technology Bangalore, Bangalore, India\\
	\email{\{shyam.sm,srinivas.vivek\}@iiitb.ac.in}}
\else
\author{}
\institute{}
\fi
\maketitle
\blfootnote{© Springer Nature Switzerland AG 2021. The final published version will be available at \url{www.springerlink.com}.}

\begin{abstract}
	Ride Hailing Services (RHS) have become a popular means of transportation, and with its popularity comes the concerns of privacy of riders and drivers.  ORide is a privacy-preserving RHS proposed at the USENIX Security Symposium 2017 and uses Somewhat Homomorphic Encryption (SHE). 
In their protocol, a rider and all drivers in a zone send their encrypted coordinates to the RHS Service Provider (SP) who computes the squared Euclidean distances between them and forwards them to the rider. The rider decrypts these and selects the optimal driver with least Euclidean distance.
\\
In this work, we demonstrate a location-harvesting attack where an honest-but-curious rider, making only a single ride request, can determine the exact coordinates of about half the number of responding drivers even when only the distance between the rider and drivers are given.
The significance of our attack lies in inferring locations of other drivers in the zone, which are not (supposed to be) revealed to the rider as per the protocol.
\\
 We validate our attack by running experiments on zones of varying sizes in arbitrarily selected big cities. Our attack is based on enumerating lattice points on a circle of sufficiently small radius and 
eliminating solutions based on conditions imposed by the application scenario. 
Finally, we propose a modification to ORide aimed at thwarting our attack and show that this modification provides sufficient driver anonymity while preserving ride matching accuracy.

\begin{keywords}
	Ride Hailing Services, Privacy and Censorship, Applied Cryptography, Lattice points.
\end{keywords}
\end{abstract}

%
\section{Introduction}
\label{sec:introduction}
Ride Hailing Services such as Uber, Lyft are becoming popular world-wide year over year. According to Pew Research \cite{pew:rhs_research}, the number of Americans who have used RHS has more than doubled since 2015.   In order to provide the service, RHS Service Providers (SP) collect upfront information about individuals desiring to use their services, which include riders and drivers who are part of the network.  In addition, details of rides offered and accepted are also collected as part of their billing and statistics gathering.  
This raises a number of privacy concerns among the individual users.   
Though the SP would, in general, keep the information secure given the need to keep its reputation high, there is nothing to prevent breach of privacy if either the provider turns malicious or if someone with access to information internal to the provider wants to mine the information for personal gain  \cite{norton_databreach}.

A ride hailing service consists of three parties, namely, the SP, a rider who has subscribed for services of the SP and a set of drivers involved in ride selection.  The SP is modeled as an honest-but-curious adversary.  We consider the threat model where the rider attempts to mount a location-harvesting attack on participating drivers.  While there are a number of solutions proposed in the last few years that preserve privacy of riders and drivers with respect to the SP, there are only a few works that look at privacy issues of drivers with respect to riders.  The work {\em Geo-locating Drivers} \cite{Zhao2019GeolocatingDA} does an analysis of features and web APIs of non-privacy preserving RHS apps which can be used to extract privacy sensitive driver data.    {\em PrivateRide} by Pham \etal \cite{Pham2017PrivateRideAP} describes how riders or other malicious outsiders posing as riders can harvest personal information of drivers for purposes of stalking, blackmailing or other malicious activities.  Apart from user-profiling, there are several instances where leakage of information regarding driver locations can lead to serious threats (refer Section \ref{sec:impact_attack}). 

One of the early privacy-preserving ride hailing services is {\em ORide} \cite{ORidePaper}.  While the primary focus of this proposal was to provide an oblivious ride-matching solution to riders while preserving the privacy of riders and drivers from the SP, it also considers
location-harvesting attacks against drivers by a malicious set of riders who create and cancel fake ride requests simultaneously from  multiple locations.   
ORide ensures the anonymity of the drivers and riders with respect to SP primarily through the use of a Somewhat Homomorphic Encryption (SHE) scheme.  
There are more recent works that also propose privacy-preserving RHS, and an overview of the works related to RHS is given in Section \ref{sec:relatedwork}.  

In ORide, the SP collects SHE encrypted coordinates of the 
drivers in the zone of the rider, homomorphically computes the Euclidean distances between the rider and drivers, 
and then sends these encrypted values to the rider. The rider then decrypts the encrypted distances, chooses the 
nearest driver and proceeds with ride establishment (the ORide protocol is recalled in Section \ref{oride_intro}).
This clearly leaks the distances of even those drivers who were {\em not} selected to offer the ride. But given that there are many possibilities for the coordinates of the driver, even if only their distance is known, one would expect that in practice the exact driver location is anonymous.
However, we show that while the protocol hides personal information of the drivers it offers only limited anonymity for the drivers' locations w.r.t.~a rider who requests a ride. 

\subsection{Our Contribution}
In this work, we show a location-harvesting attack on the drivers in the ORide protocol.
Along with the privacy for riders and drivers with respect to the SP, ORide also claims that its design  offers location privacy for drivers with respect to riders by preventing location-harvesting attacks. 
This is done using deposit tokens and permutation of driver indices for each ride request, which prevents a malicious rider from making fake ride requests and triangulating locations of all drivers in the zone \cite[\S8]{ORidePaper}. 
We show in Section \ref{analysis} that even an honest-but-curious rider, with only one ride request and response, can recover the exact coordinates of about half the number of drivers who respond to her ride request. Such an attack is not easy for the  SP to detect unlike attacks that involve simultaneous ride requests and cancellations.  We remark here that except driver location information, no personal driver information is revealed in our attack.
Nonetheless, in Section \ref{sec:impact_attack} we discuss practical scenarios where revealing only
the drivers locations (without their identities) can be harmful.


Our attack is motivated by the classical 
\textit{Gauss' circle problem} \cite[Ch. 9]{takloo2018pythagorean}. ORide uses a map-projection system such as UTM \cite{UTMGrid} to work
with planar integer coordinates. Recovering the integer coordinates $(X_d, Y_d)$ of the driver 
by the rider reduces to solving  $(X_r-X_d)^2 + (Y_r-Y_d)^2 = N$, where $X_r,Y_r,N$ are non-negative integers 
that are known to the rider. Relabelling this equation as
$x^2 + y^2 = N$ results in a variant of the Gauss' circle problem. Since $N$ is sufficiently small, it is feasible to enumerate all the lattice points (i.e., 
points with integer coordinates) on the circle of radius $\sqrt{N}$. In our case, since $N$ always 
corresponds to the case where a solution is known to exist, we experimentally observe that 
the number of solutions to be about $20$ on average (over our choice of zones in Table \ref{tab1}). Then we 
use the following ideas to further eliminate the potential solutions: (i) the driver coordinates must be in the same zone
as that of the rider, (ii) the driver is typically expected to be at a motorable location such as road 
though the rider can book the ride from anywhere. This allows us to eliminate most of the 
possibilities (see Algorithm \ref{algo:cryptanalysis} in Section \ref{sec:cryptanalysis}) and reduce the number of solutions from 20 to about 2 on  average. 
In Section \ref{practical_implementation}, we validate our attack by running experiments over 
zones of different sizes for four arbitrarily chosen big cities, and show that a rider can determine the 
\textit{exact} locations of 45\% of the responding drivers (see Table \ref{tab1}). Our attacks take an
average only 2 seconds per driver on a commodity laptop.
We stress that we are not only using the geographical information to eliminate locations, but also the fact that all coordinates are encoded as integers and hence there are only a handful of locations to enumerate on the circle in the first place. 
Our attack exploits an inherent property of SHE schemes -- namely the requirement of integer-like encoded inputs for \textit{exact} arithmetic \cite{SAC:CSVW16}. We also believe that the abstraction of our attack as enumerating lattice points on a circle (and also our extension to other distance metrics in Appendix \ref{further_attacks}) is generic and will motivate similar exploits in other privacy preserving solutions that use SHE.


In Section \ref{sec:fix}, we propose a modification to the ORide protocol which serves as a solution to overcome our attack. Here the driver obfuscates her location by choosing random coordinates within a certain distance $R$ from her original location. Now the rider receives Euclidean distances that are homomorphically computed between her and the driver's anonymized location. Accordingly we modify the rider's attack from Section \ref{sec:cryptanalysis} to account for the fact that these anonymized coordinates (which are represented by different lattice solutions) may not lie on road. However the anonymized coordinates will definitely have a road within proximity $R$, since the driver was originally on road. Through experiments we analyze this new attack on the proposed modification and evaluate its effect on driver anonymity and accuracy (refer Table \ref{table:anony-accuracy}).
The optimal driver chosen in this case (based on least Euclidean distance between the rider and anonymized drivers'  locations) is sufficiently close to the time-wise closest driver (who takes the least time to arrive at rider's location). Our solution is therefore viable in practice and is successful in preserving driver anonymity.


In Appendix \ref{further_attacks}, we investigate possible alternate modifications to ORide in an attempt to mitigate our attack. However we show that these non-trivial techniques are eventually vulnerable to the same attack. 

In Section \ref{sec:relatedwork}, we discuss related works on privacy-preserving RHS and also briefly discuss the applicability of our attacks to these works.   We also discuss other techniques
that are available in the literature for location obfuscation.

\section{Analysis of ORide Protocol}
\label{sec:cryptanalysis}
In this section we briefly recall the ORide protocol of  \cite{ORidePaper}, followed by a security analysis of the protocol at the rider's end.
We then describe our attack that would allow a rider to predict a driver's location with good accuracy, and present the results of practical experiments. 

\subsection{ORide : A Privacy-Preserving Ride Hailing Service}

\label{oride_intro}

As mentioned in Section \ref{sec:introduction}, ORide is a privacy-preserving ride hailing service
that uses an SHE scheme to match riders with drivers. 
In the process, identities and locations of drivers and riders are not revealed to the SP.  
The protocol provides accountability for SP and law-enforcement agencies in case of a malicious driver or rider.  
It also supports convenience features like automatic payment and reputation-rating of drivers/riders.  
In short, it is a complete and practical solution along with novel methods  that help keep the identity of drivers and riders oblivious to the SP, together with accountability and convenience.
The experiments done in their paper use real datasets consisting of taxi rides in New York city \cite{OrideNYCDataset}. 
Their instantiation provides 112-bit security, based on the FV SHE scheme \cite{EPRINT:FanVer12} which relies on the hardness of the Ring Learning With Errors (RLWE) problem.

We give below a high-level overview of the ORide protocol relevant to our attack.
(For more details, the reader is referred to the original paper).
The registered drivers periodically advertise their geographical zones to the SP.  
These zones are predefined by the SP and available to drivers and riders. The size of a zone
is chosen in such a way that there are sufficiently many riders and drivers to ensure
anonymity while maintaining the efficiency of ride-matching.
 When a rider wishes to hail a ride, she generates an ephemeral FV public/private key-pair $(p_k, s_k)$. 
She encrypts her planar coordinates using this key and sends it to the SP along with $p_k$ and her zone $\mathcal{Z}$.
SP broadcasts the public key $p_k$ received from the rider to each driver in $\mathcal{Z}$.
The $i$\textsuperscript{th} driver $D_i$ encrypts her planar coordinates using $p_k$ and sends it to SP.
SP homomorphically computes the squared values of the Euclidean distances between each driver and the rider in parallel, and sends the encrypted result to the rider. 
The rider decrypts the ciphertext sent by SP to obtain the squared Euclidean distance to each driver $D_i$. 
She then selects the driver with smallest squared Euclidean distance and then notifies the SP of the selected driver. 
This selected driver is in turn notified by the SP. 
As part of the ride establishment protocol a secure channel is then established between the rider and the driver. They then proceed to service the ride request as per the protocol. Further steps, although important, are not relevant to our work and, hence, we do not mention them here.

\subsection{ORide : Threat Model}

\label{oride_threat_model}

The threat model considered in ORide is that of an honest-but-curious SP, whereas the drivers and riders are active adversaries who do not collude with the SP.    We consider the same adversarial model in this paper as well.
All the plaintext information is encoded as integer polynomials before encrypting with the  FV SHE scheme.
In ORide, the apps on the drivers and riders use a map-projection system such as UTM \cite{UTMGrid} to 
convert pairs of floating-point latitudes and longitudes to planar integer coordinates.  
Drivers use third-party services like Google Maps or TomTom for navigation.

\subsection{Attack: Predicting Driver Locations}

\label{analysis}

\begin{algorithm}[h!]
\small
\SetAlgoLined\DontPrintSemicolon

\SetKwFunction{algoOne}{Predict_Driver}
\SetKwProg{myalgoOne}{Procedure}{ : }{}

\SetKwInOut{Input}{Input}\SetKwInOut{Output}{Output}
\Input{ The rider's zone $\mathcal{Z}$, number of drivers $n$ inside $\mathcal{Z}$, rider's coordinates $(X_r, Y_r)$, Euclidean distances $d_i$ between the rider and driver $D_i$ ($\forall i=1,\cdots ,n$)}
\Output{ For each driver $D_i$, $\mathcal{S}'_{i}$ denotes the prediction set made for the location of $D_i$ by the rider}

\myalgoOne{\algoOne{$\mathcal{Z}, n, (X_r, Y_r), \{d_i\}_{i=1}^{n}$}} {

\vspace{5pt}

$avg = 0$;  
$exact = 0$ 

\For{\textnormal{each driver} $D_{i}$}{
\textbf{Receive}: $d_i$ from SP \\

$\mathcal{S}_i = \phi$ \MyComment{Store unique lattice points}\\
	\For{$x = 0$ \KwTo $\lfloor \sqrt{d_i} \rfloor$}{
	$y = \sqrt{d_i - x^2}$ \\
	\If{$y$ is an integer} { 
		 $T = \{\ (x, y), (-x, y), (-x, -y),(x, -y),$\\\hfill $(y, x), (-y, x), (-y, -x), (y, -x) \ \}$ \\
		 \For{$(x', y') \in T$}{
			\MyComment{Compute predicted location for $D_i$} \\
			$X_d = X_r + x'$; 
			$Y_d = Y_r + y'$ \\
			\If{($X_d, Y_d$)\textnormal{ is inside $\mathcal{Z}$}}{
		$\mathcal{S}_i \coloneqq \mathcal{S}_i \cup \{\ (x, y)\ \}$
		}
		 }			 
	}
	}
$\mathcal{S}'_i = \phi$  \MyComment{Filtered lattice points on road} \\
	\For{$(X_d, Y_d) \in \mathcal{S}_i$}{
	\MyComment{Use Google Maps  API to check if the coordinates lie on road} \\
	${(x, y)}_{road} = \mathsf{RoadAPI}(\ (X_d, Y_d) \ )$ \\ \vspace{5pt}
	\If {$\textnormal{distance between } (X_d, Y_d) \textnormal{ and } {(x,y)}_{road} <= 3 \textnormal{ metres }$ }{

	\vspace{3pt}
	$\mathcal{S}'_i \coloneqq \mathcal{S}'_i \cup \{\ (X_d, Y_d)\ \}$
	}
	}
\MyComment{$\|\mathcal{S}'_i\|$ is the number of locations that the rider has predicted for $D_i$} \\
$avg \coloneqq avg + \|\mathcal{S}'_i\|$ \\
\If{$\|\mathcal{S}'_i\|\ == 1$}{
	$exact += 1$ 	\MyComment{Exact driver loc predictions} \\
	}
}
$avg \coloneqq avg \ / \ n$;\  
$exact \coloneqq exact \ / n \times 100$ \\ 
\textbf{Output}: $exact$, $avg$, $ \mathcal{S}'_i $ 
}

\caption{Location-harvesting Attack on ORide}
\label{algo:cryptanalysis}
\end{algorithm}

We now analyze the ORide protocol from the rider's end. 
For ease of explanation, the rest of our paper shall refer to the squared Euclidean distance between two 
points as simply the Euclidean distance. 
In the ORide protocol, before a rider finally chooses the closest driver, she is given a list of Euclidean 
distances corresponding to drivers in her zone.
In this case, the rider only gets to know the Euclidean distance to each driver, and not the driver's exact coordinates. 
Mathematically, this would mean that there are infinite possibilities for the driver's location on the 
circumference of a circle defined from the rider's perspective. 

On the contrary, we 
show that the driver's Euclidean distance allows the rider to identify the actual location of a driver with good probability.
We show that by identifying road networks on a live map (using Google Maps API \cite{googleroadapi}), along with the fact that ORide uses integer 
coordinates, the number of possible driver locations from the rider's perspective can be reduced 
significantly, to around $2$ locations on average. 

\textit{Remark}. 
While we make use of the fact that ORide uses integers coordinates, our attack would also work for fixed-
point encoding of the coordinates. This is because the current (exact) techniques for fixed-point 
encodings for RLWE-based SHE schemes essentially use the scaled-integer representation \cite{SAC:CSVW16}. 

Before we proceed with our analysis, we make the assumption that when the rider requests a ride and when 
each driver in the zone sends her encrypted coordinates to SP, the drivers are on road (since we use Google Maps API in our experiments, these include city roads, parking lot roads and many other categories, as specified by the definition of a \emph{road segment} by Google Maps \cite{googleroadapi}).
This assumption is reasonable since a vast majority of the active drivers at any point in time constantly move around the city looking for potential rides or about to finish serving another ride.

When we say that a driver's coordinates lie on road, we mean that the coordinates lie within the borders 
of the road. The current standards for lane width in the United States recommends that each lane is 3 metres wide on average \cite{roadwidth}. Since many roads within a city consist of 2 lanes, we assume that a pair of coordinates lie on road if the location is within 3 metres from the centre of a road (neighborhoods in many cities around the world consist mostly of 2 lane roads, so our experiments give a fairly accurate idea of location recovery probabilities). We stress that the drivers can be anywhere in the zone on any road and our experiments indeed follow this distribution.

\textit{Rider's attack.}
A rider performs the following attack to obtain a set of possible locations for a driver. At the time of ride request, let the rider coordinates be $(X_r, Y_r)$ and the driver coordinates be $(X_d, Y_d)$, which the rider does not know. Let the rider's zone be denoted by $\mathcal{Z}$. SP receives the encrypted values of $X_r, X_d, Y_r, Y_d$, then homomorphically computes the Euclidean distance in encrypted form, and the rider decrypts this to obtain $N = (X_r-X_d)^2 + (Y_r-Y_d)^2 $.
If $N$ is not too large (refer to Section \ref{practical_implementation} for a concrete discussion on bounds for $N$), the rider can efficiently find all integer solutions to the equation $x^2 + y^2 = N$. 
The rider could use an $O(\sqrt{N}~)$ algorithm to accomplish this: keep a solution-set, and for every 
integer $x' \in [0, \lfloor{\sqrt{N}}~\rfloor]$, compute $y' = (N - x'{}^{2})^{1/2}$. 
If $y'$ is an integer, add the coordinates $(x', y')$, $ (x', -y')$, $  (-x', y')$, $  (-x', -y')$, $  (y', x')$, $  (y', -x')$, $  (-y', x')$ and $(-y', -x') $ into this set. 

Now, rider maintains a set $\mathcal{S}$ containing the possible driver locations. For each integral solution $x_i, y_i$ satisfying $x_i^2+y_i^2=N$, the rider identifies potential driver coordinates as 
$ (X'_{d, i} = X_r + x_i, Y'_{d, i} = Y_r + y_i)$
and adds $(X'_{d, i}\,, Y'_{d, i})$ to $\mathcal{S}$ if $(X'_{d, i}\,, Y'_{d, i})$ is inside $\mathcal{Z}$.


Once the rider obtains these possible driver coordinates in $\mathcal{S}$, she checks whether each solution lies on road. (Google Maps Road API \cite{googleroadapi} can be used to achieve this). 
The rider now obtains a filtered set of coordinates $\mathcal{S^'} \subseteq \mathcal{S}$ that are inside $\mathcal{Z}$, and also lie on a road. 
Note that since the actual coordinates of the driver also satisfy these conditions, it is \emph{always} present in this set.
The cardinality of $\mathcal{S}'$ would denote the number of predicted locations for a driver. If this cardinality is exactly one, then the rider has successfully predicted the driver's exact location.
Our  attack is summarized in Algorithm \ref{algo:cryptanalysis}.   
The algorithm takes as input the Euclidean distances of drivers in the zone and, for each driver, outputs the set 
of filtered coordinates $\mathcal{S}'$.  It outputs the number of predicted locations \textit{avg}, averaged over all drivers.  Finally, it outputs \textit{exact}
which denotes the number of drivers for whom exactly one location is predicted.

\begin{figure}
\centering
\includegraphics[,width=220pt]{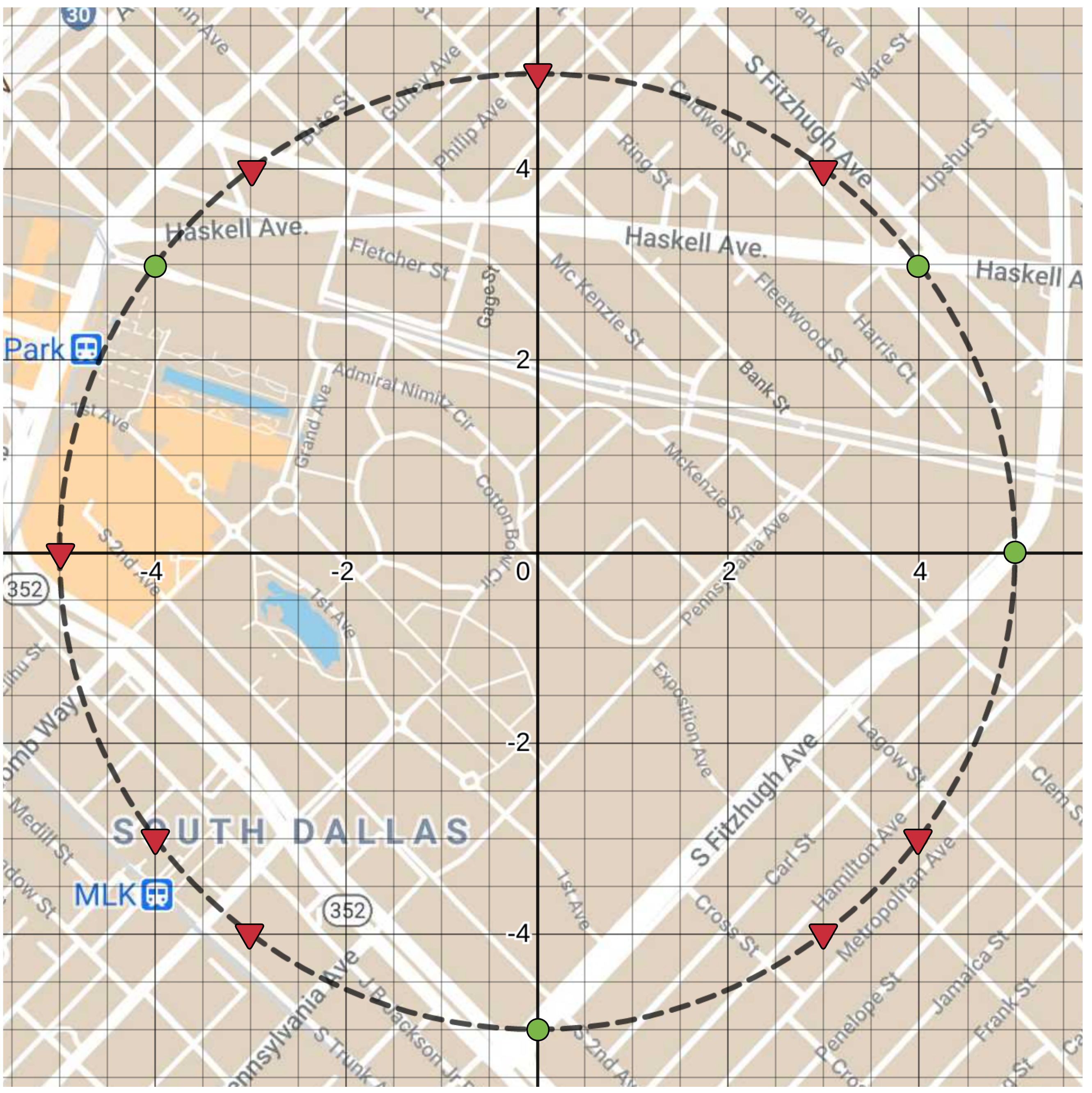}
\caption{Illustrative example of location prediction of a single driver by a rider.} \label{fig:map-example}
\end{figure}

We present an illustrative example in Figure \ref{fig:map-example}. Consider a large zone in Dallas, USA, with a cartesian grid embedded over the road view of the map. Consider a driver and a rider pair inside this zone. The rider is said to be located at the origin, and let the driver be at coordinates $ (4,3) $ (which agrees with our assumption that drivers lie on road). The rider is given the Euclidean distance $ 25 $ to this driver. She then obtains all lattice points lying on this circle: $ \mathcal{S} = \{ (\pm 3, \pm 4), (\pm 4, \pm 3), (\pm 5, 0), (0, \pm 5) \} $. Out of these, the rider filters out coordinates that lie on road (shown as green dots in Figure \ref{fig:map-example}) to obtain $ \mathcal{S'} = \{ (-4, 3), (4, 3), (5, 0), (0, -5) \} $. Note that the driver's actual coordinates belong to $ \mathcal{S'}$. Note also that if the rider's location, her zone and the Euclidean distance to this driver was given as input to Algorithm \ref{algo:cryptanalysis}, we would receive as outputs $ \mathcal{S}' = \{ (-4, 3), (4, 3), (5, 0), (0, -5) \}, avg = 4, exact = 0 $.

\textit{Remark}.
In the scenario of ORide, the rider's zone usually consists of multiple drivers. 
Note that in Algorithm \ref{algo:cryptanalysis}, when calculating the possible coordinates $\mathcal{S}'_i$ for the $i$\textsuperscript{th} driver $D_i$, the analysis that a rider performs for one driver is \emph{independent} of the analysis for other drivers. 
Therefore, the averaged results for multiple drivers in one execution of the attack 
(if the driver locations are randomly and independently sampled subject to above conditions)
is equivalent to the averaged results over multiple executions of the attack in the case of only a  
single driver present inside the zone.

Our implementation of the attack will therefore consider one driver inside the zone, and average the results over multiple experiments, considering randomly chosen rider and driver locations each time. 

\subsection{Implementation of Our Attack}

\label{practical_implementation}

Using Google Maps API for Python \cite{googleroadapi_python}, we performed experiments to validate our attack across four arbitrarily chosen cities: New York city, Dallas, Los Angeles and London 
\footnote{The code for our attack presented in Section \ref{sec:cryptanalysis} can be accessed at \href{https://github.com/deepakkavoor/rhs-attack}{https://github.com/deepakkavoor/rhs-attack}}.
We ran our experiments over zones of sizes $A = $ \{ 1 km\textsuperscript{2}, 4 km\textsuperscript{2}, 9 km\textsuperscript{2}, 25 km\textsuperscript{2}, 100 km\textsuperscript{2}, 400 km\textsuperscript{2}, 900 km\textsuperscript{2} \} (with the exception of New York city due to its geography having multiple smaller discontinguous areas). 
For each city, and for each zone size $a \in A$, we performed 30 experiments. In each experiment, a random square zone $\mathcal{Z}_a$ of area equal to $a$, was chosen.
We chose a random latitude-longitude pair inside $\mathcal{Z}_a$ for the rider in this zone. 
For a driver, we similarly chose a random latitude-longitude pair inside $\mathcal{Z}_{a}$ that was on road (Google Maps Road API was used to accomplish this).

These coordinates were converted to UTM coordinates using the \emph{utm} library for Python \cite{utmPython}.
The Euclidean distance between these UTM coordinates was made available to the rider.
Finally, we obtained the driver's filtered set of probable locations as described in Algorithm \ref{algo:cryptanalysis}. After obtaining the predicted driver coordinates, we averaged the number of such predicted locations over multiple experiments. We also counted the percentage of experiments in which the rider predicted exactly one location. 
With these considerations, our results for varying grid sizes and cities are shown in Table \ref{tab1}. 

\begin{table*}
	\small
	\renewcommand{\arraystretch}{1.3}
	\caption{The rider's prediction based on Algorithm \ref{algo:cryptanalysis}, averaged over 30 experiments.}
	\label{tab1}
	\centering
	\renewcommand*{\arraystretch}{1.1}
	\begin{tabular}{c|c c c c|c c c c}
		Zone Size& 
		\multicolumn{4}{p{5cm}}{\centering Number of possible driver locations \\ (Output $avg$ of Algorithm \ref{algo:cryptanalysis})} &
		\multicolumn{4}{|p{5cm}}{\centering Exact driver locations prediction \\ (Output $exact$ of Algorithm \ref{algo:cryptanalysis})} \\[0.2em]
		\cline{2-9}
		(km\textsuperscript{2}) & New York & Dallas & Los Angeles & London & New York & Dallas & Los Angeles & London  \\
		\hline 
		1   & 2.6 & 1.8 & 2.5 & 1.6 & 32\% & 52\% & 32\% & 60\% \\
		2   & 2.4 & 1.6 & 2.1 & 1.6 & 48\% & 52\% & 36\% & 68\% \\
		4   & 2.0 & 2.0 & 1.8 & 1.7 & 44\% & 44\% & 52\% & 56\% \\
		9   & 2.7 & 1.9 & 2.1 & 2.3 & 38\% & 56\% & 48\% & 40\% \\
		25  & 2.6 & 2.2 & 2.4 & 2.1 & 36\% & 44\% & 36\% & 48\% \\
		100 & 2.7 & 2.1 & 2.2 & 1.8 & 32\% & 44\% & 44\% & 56\% \\
		400 & 2.3 & 2.1 & 2.5 & 1.8 & 36\% & 48\% & 28\% & 56\% \\
		900 &  -- & 2.8 & 3.1 & 2.4 &  --  & 40\% & 24\% & 40\% \\
	\end{tabular}
\end{table*}

\subsubsection{Timings} 
Our experiments were performed on an Intel Core i5-8250U CPU @ 1.60 GHz with 8 GB RAM running Ubuntu 18.04.4 LTS.
On an average, one experiment (as described above) took 2 seconds for each driver, showing that our attack is indeed efficient, thus allowing a rider to practically obtain any driver's coordinates with good confidence.

\subsubsection{Interpretation of values in Table \ref{tab1}}
Our experiments showed that the average number of solutions to $x^2+y^2=N$  over all the aforementioned zone sizes was 20. When we filter these solutions based on whether they lie inside a zone and on road, the average possible driver coordinates were 2 in number (as indicated by the value \emph{avg} in Table \ref{tab1}), which is a significant reduction. Although it may seem that Euclidean distance gives fair anonymity to driver coordinates, our attack shows that in practice, this is not the case, and a rider can indeed find the driver's location with good probability. 
We also note from the average value of \emph{exact} in Table \ref{tab1} that the rider can predict the driver's exact location around 45\% of the times.

Note that in each city, as zones get bigger, the number of lattice solutions and filtered coordinates tend to increase leading to higher \emph{avg}. More lattice solutions imply that the event when a rider predicts exactly one location for a driver is rare, thus decreasing the value of \emph{exact}. This trend can be verified from  Table \ref{tab1}.


\subsubsection{Anonymity Sets}

In ORide, when the rider makes a request, she sends her zone identity to SP, and the SP now knows which zone the rider is in. 
This zone could contain the rider's home/work address. 
As pointed out in the ORide paper \cite{ORidePaper}, SP might be able to guess the identities of the riders if this pick-up zone had a limited number of ride activities, and a limited number of riders (as an extreme example, a zone where only one rider lives).
Therefore, ORide defines zones in such a way that each zone has at least a large minimum number of ride requests per day. 
This large minimum is referred to by them as the \emph{anonymity-set} size.
The choice of size of these zones is left to the SP, based on balancing the communication bandwidth requirements and sizes of anonymity sets in those zones.
(A very high anonymity set would mean that the demand for rides in that zone is high, leading to longer ride matching times and higher bandwidth usage).

We justify our choice of choosing zones of sizes $a \in A$ for our experiments:
\begin{compactlist}

\item In a densely populated city like New York City (population density\footnote{\href{https://worldpopulationreview.com/us-cities}{https://worldpopulationreview.com/us-cities}} 11,084 persons/km\textsuperscript{2}), where more people tend to use ride hailing services, a smaller zone size would suffice to achieve the required anonymity-set size.
In a sparse city like Dallas (population density 1,590 persons/km\textsuperscript{2}), where fewer ride-hailing activities occur, these zones would have to be bigger in size to achieve the same anonymity for riders.
Taking into consideration the different possible zone sizes in both densely populated and sparse cities, the experiments validate our attack in zones of areas ranging from 1~km\textsuperscript{2} to 900~km\textsuperscript{2}.

\item We analyzed the NYC Uber-Dataset \cite{NYCDataset} for May 2019, and deduced that the demand for taxi rides was very high in Manhattan compared to the other boroughs of NYC. We chose May since this month had one of the highest ride requests for Uber in 2019.
Based on this, we followed the zone demarcation that was proposed by ORide: each Census Tract (CT) \cite{NYCensusTract} in Manhattan is considered as one zone. The boroughs of Queens and Bronx are merged into one zone, and the boroughs of Brooklyn and Staten Island are merged into one zone.
The size of each CT in Manhattan varies between 1 km\textsuperscript{2} and 4 km\textsuperscript{2} that correspond to zone size of higher activity.  Since the boroughs other than Manhattan have lesser activity, these zones are expected to have a larger area. Indeed, the combined area of Queens and Bronx is around 390 km\textsuperscript{2}, and the combined area of Brooklyn and Staten Island is around 330 km\textsuperscript{2}. Since this is the primary zone demarcation proposed by the authors of ORide, we found it reasonable to include these ranges of areas for our experiments in Table \ref{tab1}.

\end{compactlist}

We next discuss few details involved in the implementation of our attack.
\begin{compactlist}
\item  Although zones can be of any geographical shape, we chose square zones for ease of choosing 
random coordinates inside its boundary, and to simplify checking whether a given coordinate lies inside the zone.

\item Latitude-longitude coordinates were converted into UTM formats using the \emph{utm} library for Python. 
On an average, this conversion  results in a difference of 0.5 metres between the original coordinate and the planar coordinate's representation.
For all practical purposes, this difference is very small, and the two coordinates can be considered to represent the same location.

\item As discussed earlier, based on the NYC Uber-Dataset and ORide's proposed demarcation, even large sparse zones that have a sufficiently big anonymity-set would rarely exceed $30\times 30=900$ km\textsuperscript{2}. 
Note that the Euclidean distance between two UTM coordinates is equal to the distance (in metres) between latitude-longitude representation of those points. 
Hence, the maximum value of $N$ for a $d \times d$ grid would be $2d^2$.  For a 
30 km $\times$ 30 km grid, $d$ would be $30,000$.  
Since there exists an $O(\sqrt{N}) = O(d)$ algorithm to compute solutions to $x^2+y^2=N$, it is indeed feasible for the rider to perform this analysis on modern computers in very less time, even for different zone structures chosen by SP.  
\end{compactlist}

\textit{Remark}.
We give a brief insight into the number of drivers inside a zone, which averages to 400. The zone demarcation proposed for New York by the authors of ORide was discussed briefly above. 
According to ride information for May 2019 in the NYC Uber-Dataset, a zone in Manhattan had at most 6,000 ride requests per day.  (We chose the month of May since it experienced the most ride-requests in the year 2019).
We make the same assumption that the authors of ORide did: the drop-off zone for a driver is her waiting zone for new ride requests. Moreover, as in ORide, we assume that the waiting time between a driver's drop-off event and her next pick-up event is at most 30 minutes. 
This would mean that during a ride-request event, the available drivers to answer this request are the ones who had a drop-off event inside that zone in the last 30 minutes since the ride-request. 
We considered the top 20 high-ride zones, and for each zone, grouped the ride requests for a day based on 30 minute intervals. 
Each 30 minute interval consisted of at most 400 drop-off events inside each zone. This would imply that when a ride request occurs at any time of the day, at most 400 drivers would be waiting in that zone to service this request. We stress that there are at most 400 drivers in all zone demarcations considered above, and as the zone size increases the density of drivers (the number of drivers available for ride request in 1 km\textsuperscript{2}) in that zone decreases.


\subsection{Impact and Consequences of our Attack}
\label{sec:impact_attack}

We have experimentally shown that our location-harvesting attack can identify the exact locations of a driver in about 45\% of the cases. Equivalently, this means that in a zone of around 400 available drivers, a ride request leaks the locations of around 180 drivers to the rider, which is a significant number. Although our attack doesn't reveal additional driver data such as user profiles, this leak of location information could still cause potential threats to drivers and ultimately affect the SP's reputation.
For instance, according to \cite{istanbul_uber}, it is claimed that non-SP taxi drivers try to identify locations of Uber vehicles and attack them. There are also reports of people using ride-hailing apps to locate and rob drivers registered to the SP \cite{thejournal}. Zhao et.~al.~\cite{Zhao2019GeolocatingDA} investigate several approaches through which information regarding drivers' locations can lead to statistical attacks and exploits. 
A potential competitor to ORide can use our attack and make queries to ORide as an honest-but-curious rider over a period of time.   It can then get to know the distribution/density of different drivers in the city without raising suspicions. This distribution could indicate regions where there is high demand for ride hailing services.
The competitor could focus on deploying their drivers and taxis in that region.

In general, ensuring privacy of the locations of drivers in the zone should be an important aspect of any privacy-preserving RHS. The authors of ORide claim that an adversary cannot obtain a snapshot of drivers' locations (say, a malicious rider who makes multiple fake ride requests with the goal of harvesting drivers' locations) \cite[\S8]{ORidePaper}, thus preserving their anonymity.
In contrast, our work refutes this claim made in ORide using just a single ride request by an honest-but-curious rider.
We think that this flaw in ORide is not merely an implementation error. The requirement of integer-encoded inputs is inherent to current SHE schemes, and this helps us obtain small number of lattice points on the circle.

\textit{Remark. }
One of our motivations behind considering privacy of driver locations is to prevent physical attacks on the drivers in a zone. It can be argued that even if locations of these drivers are not revealed, a malicious rider can request a ride in an honest manner, and attack the chosen driver when she arrives at the pick-up location. 
While this is true, and in fact, at the end of any ride hailing service (whether it preserves privacy or not), the selected driver \textit{must} arrive at the rider’s pick-up location and so, this attack scenario is nearly impossible to thwart. However, in almost all apps (including ORide), the identity information of the rider (name, phone number etc.) is revealed to the driver selected for the ride (and vice-versa) which act as a 
deterrent for such attacks. 
The possibility of fake accounts can be eliminated by enforcing identity verification at the time of registration. This does not violate rider privacy since ORide assigns anonymous tokens to all entities during the ride request process. 
Using our attack from Section \ref{analysis}, a rider can get to know even the locations of \textit{non-selected} drivers in the zone, namely the other drivers who have no information about the identity of the rider. 
Finally, we remark that if the driver has moved from her last revealed location or does not have a logo/sticker on her car (advertising that she belongs to a particular RHS) then it would be hard for the rider to carry
out physical attacks.

\section{Mitigation of our Attack}
\label{sec:fix}

We propose a solution where the driver can thwart our attack by anonymizing her location. Each driver could choose a random coordinate within a circle of fixed radius around herself, encrypt and send these random coordinates to the SP instead of her original coordinates. We show that this modification to ORide provides sufficient anonymity while preserving ride matching accuracy, and is therefore a reasonable solution to mitigate our attack.   The idea of adding noise to geographical data to preserve privacy has been addressed in
several works in the literature. In Section \ref{sec:relatedwork}, we discuss some of these techniques and their relevance to our setting.
We analyze the effect of this technique on driver anonymity and provide concrete values for ride matching accuracy through experiments using Google Maps API.

\textit{Remark.} In Appendix \ref{further_attacks}, we discuss other ideas that may intuitively seem to thwart our attack. However, we show that those modifications are vulnerable to our attack from Section \ref{sec:cryptanalysis} and hence do not preserve driver anonymity.

\subsection{Anonymizing Driver Locations}
\label{sec:fix_idea}

By anonymizing her location, each driver may try to preserve the privacy of her location with respect to a rider.
Let each driver $D_i$ (at coordinates $L_i$) choose a circle of radius $R$ centered at her location (where $R$ is publicly known), and pick a random UTM  coordinates $L_i'$ inside this circle. The driver encrypts $L_i'$  (instead of $L_i$ as suggested by ORide) and sends it to SP. We refer to $L_i'$ as the anonymized driver coordinates.

As per the original attack in Section \ref{sec:cryptanalysis}, the rider obtains a Euclidean distance $N_i$ and enumerates all lattice points that correspond to this distance. Due to the changes described above, these lattice points need not correspond to possible driver coordinates. They instead represent possible \textit{anonymized} driver coordinates. 
The rider would have next proceeded to filter each lattice point based on whether it is on road or not. But a lattice point which represents an anonymized driver location may not lie a point on road, although the original driver did. Filtering in this way would lead to erroneous conclusions by the rider, and she may throw out a lattice point that actually corresponds to the driver location. 

Observe that within distance $R$ of anonymized driver coordinates, there will always lie a road (since the original driver was on road). We modify the rider's attack accordingly to cope with this fix. Suppose there was a lattice point discovered by the rider. Within a circle of radius $R$ centered at this point, if there were no roads at all (for instance a lattice point that was in the middle of a park) the rider can then conclude that this point is not the driver's anonymized coordinates. So, the best option that a rider has (to improve her attack against this obfuscation technique) is to filter each lattice point based on whether there is a road within distance $R$ of that point. 
As we see in Section \ref{sec:effect-on-anonymity}, the possibility that a lattice point is filtered out in a dense city is low if we choose an appropriate value of $R$. This prevents a rider from eliminating many lattice points thus improving driver anonymity. Moreover, this technique preserves accuracy when compared to ORide as we show next.



  
\subsection{Anonymity of Drivers with respect to Rider}
\label{sec:effect-on-anonymity}

The value of $R$ is public and should be decided by the SP, who can in fact implement the end-user application in such a way that the driver's device locally computes $R$ based on the current location of the driver. If the driver's local device senses that she is in a densely populated city (and thus there are many roads within close proximity of an arbitrary point in that region of the city), a smaller $R$ can be chosen. On the other hand, if the device understands that the driver is in a location where there are very few roads within distance $R$ from an arbitrary point in that region, a larger $R$ is chosen (for instance, in a sparsely populated city with low road density). This choice of $R$ based on the concentration of roads around the driver is motivated by the modification to rider's attack discussed at the end of  Section \ref{sec:fix_idea} (the rider's attack now tries to filter lattice points based on the availability of roads within distance $R$ from each lattice point solution).
We consider the number of (anonymized) driver locations predicted by a rider as a measure of anonymity for that driver. This depends on the number of lattice solutions for the Euclidean distance between (anonymized) driver location and rider. Along with this, it also depends on the number of solutions that the rider can further filter based on availability of roads within distance $R$ from each solution. We expect anonymity to increase with $R$ due to higher probability of finding a road within distance $R$ of any location.

Similar to the setup in Section \ref{practical_implementation}, in the following discussion we average results over 25 experiments where each experiment chooses a random zone of size 4 km\textsuperscript{2} in the mentioned city (along with random coordinates for a rider and driver) and runs the modified rider's attack for filtering coordinates.
For a small value of $R$ such as 10 m, any coordinate within distance $R$ of some point is practically the same location.
We experimentally observed that the average anonymity for a driver in Los Angeles was around 3, which is close to what we observe in the original attack (see Table \ref{tab1}). Hence small values of $R$ should not be chosen since they offer low anonymity. 
In a densely populated city such as Los Angeles, most locations within the city are expected to have roads within reasonable distance. For $R=$~50~m  we observed that the area surrounding \textit{most} lattice points in Los Angeles had at least one road within 50~m. From a rider's perspective, this would mean that most lattice solutions obtained by her are possible choices for the anonymized driver coordinates. Experiments showed that the average number of filtered lattice points when $R=50$~m was 14 (meaning the rider has 14 possible \textit{anonymized} locations of a driver). This provides sufficient anonymity to a driver in practice, since the probability of correctly predicting a driver's anonymized coordinate is only $1/14$.   This is certainly an improvement compared to $1/1.8$ for ORide (Table \ref{tab1}).
Considering Dallas, a city with relatively sparse road density, we observed that a significant number of locations did not have roads within 50~m, and this allowed the rider to filter out many possible lattice solutions. Our experiments suggest that choosing $R=150$ m prevents the rider from doing so and offers sufficient anonymity, which averaged around 16.

\textit{Remark.} 
As seen above, the rider ends up with around 15 equally probable solutions for the driver's obfuscated
location.  This means that even if the rider applies clustering algorithms over multiple queries to eliminate noise and tries to find the actual driver’s location, there will still be many equally probable locations for the driver ($>10$) thus providing high anonymity.

\subsection{Accuracy of Ride-matching}
When a driver chooses random coordinates within distance $R$ instead of her own location, the Euclidean distance is now computed between the rider's location and the \textit{anonymized} driver location. 

Among all drivers in the zone, suppose an optimal driver is chosen according to some metric $M$. For example, if $M$ represents Euclidean distance, the optimal driver is the one with least Euclidean distance from her location to the rider in the case of ORide, and the one with least Euclidean distance from her anonymized location to the rider in the case of our modified solution. Let $t_M$ be the time taken for this optimal driver to reach rider. Let $t_T$ be the minimum time taken among all drivers in the zone to reach rider (corresponding to the time-wise closest driver).
We evaluate the accuracy of metric $M$ as the percentage of experiments in which $|t_M-t_T|$ is less than or equal to 1 minute (in practice it is okay for the rider to wait another extra minute compared to the time-wise closest driver). Google Maps API was used to determine the time taken for a driver to reach the rider.

We chose zones of 
varying sizes in Los Angeles and Dallas. In each experiment a random rider and 400 drivers were chosen in each zone.
We compared the accuracy of Euclidean metric for $R$ = 50~m and $R$ = 150~m in both scenarios -- when used in the context of ORide (computed between the rider and driver's actual location) and when used in the fix to our attack (computed between the rider and driver's anonymized location). As discussed previously, we chose $R=50$~m for Los Angeles and $R=150$~m for Dallas, respectively, to ensure sufficient anonymity. Moreover, the sizes of zones are chosen to be smaller in Los Angeles (refer Section \ref{practical_implementation}) and larger in Dallas. 
The inferred accuracies were averaged over 25 experiments (refer Table \ref{table:anony-accuracy}). We see that our solution indeed provides sufficient driver anonymity with respect to rider while preserving accuracy of ride matching compared to ORide. 

Choosing large $R$ to achieve greater anonymity in a small zone (where driver density is high) leads to loss of accuracy. This seems intuitively correct, since having a large anonymity radius in a small zone with high driver density greatly changes the ordering of drivers based on Euclidean distances. To concretely verify this, we used a similar setup described above and observed that with $R$ = 150~m and a 4 km\textsuperscript{2} zone size in Los Angeles the accuracy of ORide was around 84\% whereas that of the modified solution was only 70\%. So, $R$ should increase with zone size both to preserve accuracy and driver anonymity (prevent filtering of lattice solutions based on availability of roads).

\begin{table}[!t]
\renewcommand{\arraystretch}{1.3}
\caption{Comparison of accuracy of selecting best driver in ORide vs. our solution (with anonymized driver locations), averaged over 25 experiments.}
\label{table:anony-accuracy}
\centering
\begin{tabular}{l|c|c|c|c}
City & Zone & Radius & ORide & Our\\
      & Size (km\textsuperscript{2}) & $R$  (m)  &  & solution\\
\hline
\multirow{3}{5em}{Los~Angeles} & 4 & 50 & 84\% & 80\% \\
& 25 & 50 & 92\% & 90\%\\
\hline
Dallas & 100 & 150 & 83\% & 83\%\\
\end{tabular}
\end{table}


%
%
%
%
\section{Related Works}
\label{sec:relatedwork}


Among providers of RHS namely Lyft, DiDi, OLA, taxify and others, Uber is one of the popular ride service providers.  An in-depth analysis of the practices followed by Uber and the impact of price-surging on passengers and drivers are done by Chen \etal \cite{Chen2015PeekingBT}. {\em The Guardian} \cite{guardianReport} reports how anonymized details of New York city taxi drivers can be used to easily convert the data to its original format to  obtain personal information.
Different threat models are widely considered in the literature, namely, a malicious driver targeting riders, and an honest-but-curious SP harvesting information about riders and drivers with the intention of selling it to other entities for advertising purposes, or with potentially malicious intentions to target high profile individuals.  Privacy of the driver is given much less attention; so much so that in a few papers the actual driver locations are revealed to the SP as well as the rider \cite{khazbak}  and \cite{raza_bride}.   As motivated in Section \ref{sec:introduction} there can be instances where a malicious rider can target drivers of a specific SP.    For example, a competitor SP can masquerade as rider to collect driver profile information or statistics to target the drivers belonging to the specific SP.  
{\em Geo-locating Drivers} by Zhao \etal\cite{Zhao2019GeolocatingDA} does a study of leakage of sensitive data, in particular, it evaluates the threat to driver information.  They show it is possible to harvest driver data by a malicious outsider SP by analyzing APIs in non-privacy preserving apps provided to drivers by Uber, Lyft and other popularly deployed SPs. \linebreak
\indent {\em  PrivateRide} by Pham \etal \cite{Pham2017PrivateRideAP} is one of the first papers to address privacy in RHS.  The location of the riders are kept hidden by means of a cloaked region, and location privacy is preserved by using cryptographically secure constructs.  Details of rider and selected driver are mutually exchanged
only after the ride request is fulfilled and when they both are in close proximity, to prevent a malicious outsider
trying to harvest driver information.
A recent work by Khazbak \etal \cite{khazbak} improves upon the solution of {\em PrivateRide} by providing obfuscation techniques (spatial and temporal cloaking), of rider locations, to achieve better results in terms of selecting the closest driver, at the cost of slightly more computational overhead.   However, the drivers' locations are revealed to the rider.  \\
\indent {\em ORide} \cite{ORidePaper} is a follow-up work by the same authors of {\em PrivateRide} that provides more robust privacy and accountability guarantees, and has been described earlier in this paper.  All the following works try to improve upon ORide by proposing different models of privacy-preserving closest driver selection by the SP.  We note here that our attack is relevant in cases where the rider gets to make a choice, and is not applicable in situations where the SP selects a single suitable driver and provides the same to the rider.
{\em pRide} by Luo \etal \cite{pRideLuo} proposes a privacy-preserving ride-matching service involving two non-colluding servers with one being the SP and the other a third-party Crypto Provider (CP).   The solution makes use of Road Network Embedding (RNE) \cite{shahabiRNE} technique to transform a road network into a higher dimensional space so that the distance computation between any two nodes in the network can be performed efficiently.  They propose two solutions, one using the Paillier cryptosystem and another using BGN cryptosystem.  The homomorphically encrypted driver and rider locations received by the SP are sent to the CP along with a random noise where it is decrypted and garbled.  The SP then uses a garbled circuit to find the closest driver to the rider and completes the ride request.  They show high accuracy in matching the closest driver while preserving the privacy of driver and rider locations.   The disadvantage of this scheme is their use of a second Crypto Server that does not collude with the SP, which may be inconvenient to realize in practice, and also the high communication cost between the two servers.   
{\em lpRide} by Yu \etal \cite{lpRideYu} improves upon {\em pRide} to perform all the homomorphic distance computation algorithms on a single SP server thus eliminating high communication cost when two servers are involved.  They use modified Paillier cryptosystem \cite{nabeelPaillier} for encrypting RNE transformed locations of rider and driver.  However, \cite{lpRideAttackSV} proposed an attack on the modified Paillier scheme used in lpRide, allowing the service provider to recover locations of all riders and drivers in the region. 
Wang \etal propose {\em TRACE} \cite{wangTrace} that uses bilinear pairing for encrypting driver and rider locations.  {\em PSRide} by Yu \etal \cite{yuPSRIde} uses Paillier cryptosystem and Yao's garbled circuit with two servers on the same lines as {\em pRide} and hence suffers from some of the disadvantages mentioned above.
\textit{EPRide} by Yu \etal \cite{epRideYu} efficiently finds the exact shortest road distance using a road network hypercube embedding. They experimentally show significant improvements in accuracy and efficiency compared to \textit{ORide} and \textit{pRide}. Xie \etal \cite{xieTifs2021} compute shortest distances using road network embeddings along with property-preserving hash functions. In doing so, they remove the need for a trusted third-party server.  \\
\indent Maouche \etal \cite{maouche_attack} propose a user re-identification attack on four different mobility 
datasets obfuscated using three different Location Preserving Privacy Mechanisms (LLPM), with one of the datasets in the RHS setting.   Their attack makes use of previously learned user profiles and tries to re-
associate the same to obfuscated data.
A number of LLPMs are available in the literature that anonymize private data.  
Differential privacy and $k$-anonymity are two popular techniques.  Differential privacy, introduced by the seminal work of Dwork \etal \cite{dwork_diff_privacy} can be applied wherever
aggregate information from several similar entities are available.   
Geo-indistinguishability by Andr\'{e}s \etal \cite{geo_indis}
adds exponentially decaying noise from a Laplace distribution around the point of interest thereby obfuscating the point of interest.  
The notion of $k$-anonymity by Sweeney 
~\cite{sweeny_k_anon} obfuscates an entity by introducing $k-1$ dummy uniformly distributed entities which
are indistinguishable by the adversary.  
In our case, the driver applies noise to her coordinates before encrypting and sending to SP.   SP homomorphically computes the Euclidean distance between the rider and each driver.   
Using this (noisy) Euclidean distance, the rider solves the Gauss circle problem and filters out solutions depending on whether they have a road in their vicinity. Finally, the rider ends up with not one, but many possible choices for the anonymized driver location. This is a combination of both differential privacy (where the driver applies noise to her location) and $k$-anonymity (the rider has many equally possible choices for the driver’s obfuscated location). 
Empirically, we see that our method of adding uniformly random noise is 
sufficient to provide high anonymity to the driver.   Also, our method of filtering out non-plausible
driver locations is based on the region's topography.  We leave
the analysis of using other obfuscation techniques to thwart our attack for future work.



%
%
\vspace{-0.1in}
\section{Conclusion}
\vspace{-0.1in}
\label{sec:conclusion}
In this paper we present an attack on a privacy-preserving RHS, ORide \cite{ORidePaper}.  We show that an honest-but-curious  rider can determine the coordinates of nearly half the number of drivers in a zone even when only the 
Euclidean distance between the rider and a driver is available to the rider. 
Our attack involves enumeration of lattice points on a circle of appropriate radius and subsequent elimination of lattice points based on geographic conditions.
Finally we propose a modification to the ORide protocol as a strategy to mitigate our attack. Here a driver anonymizes her location by choosing a random coordinate within a circle of certain radius around herself. We show through concrete experiments that this technique preserves driver anonymity and accuracy of ride matching.

Although protocols may seem secure in theory, there may arise several complications and vulnerabilities when they are deployed practically, as demonstrated by our attack in Section \ref{sec:cryptanalysis}. In the future it will be interesting to experimentally investigate the notion of driver privacy with respect to both the SP and rider in more recent works following ORide (\textit{lpRide}, \cite{lpRideYu}, \textit{pRide} \cite{pRideLuo}).
\vspace{-0.1in}
\subsubsection*{Acknowledgements.}
The authors would like to thank Sonata Software Limited, Bengaluru, India for funding this work.  
We also
thank the anonymous reviewers of ACM CCS2020, USENIX Security 2021 and SAC 2021 for their
valuable comments and suggestions.

\bibliographystyle{splncs04}
\bibliography{abbrev0,crypto,morerefs}

\appendix
\vspace{-0.15in}
\section{Appendix : Further Attacks}
\label{further_attacks}
We look at potential ways in which our attack can be thwarted and analyze their efficacy.  In the first scenario, in order to obfuscate driver locations, the SP homomorphically adds  noise to driver distances before sending them to the rider. 
 For this case, we show that a rider can still break anonymity by recovering the original distances between the rider and the drivers.  
 In the second scenario, the SP uses $p$-norm metric instead of the Euclidean distance and we show that our attack also extends to this case.
Note that increasing zone sizes is not a countermeasure to our attack. As discussed in Section \ref{practical_implementation}, zone sizes should be small enough (less than 1000 km\textsuperscript{2} in practice) to ensure efficient ride-matching times and lower bandwidth costs.
 

\vspace{-0.1in}
\subsection{Homomorphic Noise Addition by SP}

In order to thwart our attack, the SP could try to obfuscate driver locations by
transforming the (encrypted squared) Euclidean distances using a random monotonic polynomial $F$ with integer coefficients and of a small degree, as suggested by Kesarwani \etal \cite{KesarwaniKNPSMM18}. 
Integer coefficients are needed for ease of representation in homomorphic computations, monotonicity is needed to maintain the sorting order of the distance inputs (so that the rider obtains the correct order upon decryption), and low polynomial degree is required for efficient homomorphic evaluation.
Let $N_i$ be the Euclidean distance between the rider and a driver $D_i$ in her zone.
The rider would get to know from the SP, for each driver $D_i$ in her zone, the values $F(N_i)$ for some random monotonic integer polynomial $F$ of low degree. Note that $F$ is unknown to the rider, but the degree $d$, range of coefficients of the polynomial $[1,2^\alpha -1]$ and range of $N_i$ ($[0,2^\beta -1]$) are publicly known. We claim that the rider can obtain the actual distance $N_i$.

\cite{MurthyV19} provides a method of recovering a monotonic integer polynomial of low degree and bounded input range when only sufficiently many outputs evaluated at integer points are provided.  We used the publicly available SageMath \cite{knnpolycode} code from the authors of \cite{MurthyV19} with parameters similar to that described in \cite{KesarwaniKNPSMM18}, namely $d = 9$, $\alpha = 32$ and $\beta = 28$. 
Next one obtains outputs $F(N_i)$ by evaluating this polynomial on the distances $N_i$.  These two steps are the same as what the SP would do (homomorphically) once it receives inputs from the rider and all drivers in a particular zone.  The $F(N_i)$ values, $d$, $\alpha$ and $\beta$ are the only values given to the SageMath code in the experiments to recover (squared) Euclidean distances to drivers for various zone sizes.  We correlated back the results of the recovery with the input distances and verified that in all cases the recovered distances matched correctly, which means that the rider can proceed with the attack mentioned in Section \ref{sec:cryptanalysis} after recovering $N_i$ values.   

\vspace{-0.1in}
\subsection{$p$-norm Metric by SP}
\label{sec:p-norm}
In order to mitigate our attack in Section \ref{sec:cryptanalysis}, the SP may try to homomorphically compute the $p$-norm (instead of Euclidean distance) of ciphertexts and send it to rider. Let $(x_R, y_R), (x_{D_i},y_{D_i})$ denote coordinates of a rider and driver $D_i$, respectively. The rider would thus obtain $N_i = |x_R-x_{D_i}|^p+|y_R-y_{D_i}|^p$ for each driver $D_i$ in her zone (the value of $p$ should not be too large to allow efficient homomorphic computations by SP).

Note that if $p$ is odd, $(x_R-x_{D_i})^p+(y_R-y_{D_i})^p$ could represent a negative value. Since ORide uses \textit{ciphertext packing} and non-Boolean circuit representation with the underlying SHE library \cite{ORidePaper}, it is very inefficient to compute the absolute value homomorphically. 
Hence, the SP would have to use only even values for $p$.
Let $p=2q$. In the rider's attack, she has to now enumerate all lattice points satisfying the equation $x^p+y^p=N_i$. Observe that if $(x,y)$ is a solution to this equation, then the lattice point $(x^q, y^q)$ is a solution to $x^2+y^2=N_i$. This implies that the solution set comprising of lattice points satisfying $x^p+y^p=N_i$ is smaller than the solution set of lattice points satisfying $x^2+y^2=N_i$. Based on our experiments on various zones and cities, we have estimated the number of lattice points satisfying $x^2+y^2=N_i$ to be around 20 (refer to Section \ref{practical_implementation}). This means that on average, the lattice points satisfying $x^p+y^p=N_i$ cannot be greater than 20 in number. The rider (similar to the rest of the attack) can then check whether each lattice point lies in the zone and on road, to reduce the number of possible predicted driver locations. In this way, our attack also applies when the SP uses $p$-norm instead of Euclidean distance. 


\end{document}